\documentclass[aip,preprint,amsmath,amssymb, 
]{revtex4-1}

\usepackage{graphicx}
\usepackage{dcolumn}
\usepackage{bm}
\usepackage{color}

\usepackage[utf8]{inputenc}
\usepackage[T1]{fontenc}
\usepackage{mathptmx}
\usepackage{etoolbox}
\usepackage{booktabs}
\usepackage{soul}
\makeatletter
\def\@email#1#2{%
 \endgroup
 \patchcmd{\titleblock@produce}
  {\frontmatter@RRAPformat}
  {\frontmatter@RRAPformat{\produce@RRAP{*#1\href{mailto:#2}{#2}}}\frontmatter@RRAPformat}
  {}{}
}%
\makeatother
\begin{document}
\preprint{AIP/123-QED}

\title[Wannier-first approach]{A Wannier-first approach for extended chiral systems}
\author{Yashpal Singh}
\email{singh5y@cmich.edu}
\affiliation{Department of Physics, Central Michigan University, Mt. Pleasant, Michigan 48859, USA}

\author{Juan E. Peralta}
\affiliation{Department of Physics, Central Michigan University, Mt. Pleasant, Michigan 48859, USA}

\author{Koblar A. Jackson}
\email{jacks1ka@cmich.edu}
\affiliation{Department of Physics, Central Michigan University, Mt. Pleasant, Michigan 48859, USA}

\author{Mark R. Pederson}
\affiliation{Department of Physics, University of Texas at El Paso, El Paso, Texas 79968, USA}
\date{\today}

\begin{abstract}
We present a real-space formulation of density functional theory for extended systems in which localized Wannier-like functions are constructed directly from localized Gaussian basis functions without explicitly computing canonical Bloch-like states during the self-consistent cycle. Building on the formalism of Pederson and Lin [Phys. Rev. B \textbf{35}, 2273 (1987)], a variational set of Wannier-like functions is generated self-consistently within a finite Wannier domain and used to construct the charge density, electrostatic potential, and per cell total energy of the extended system. The occupied space can be determined entirely from the Wannier-like functions. Electronic band structures can be recovered in a post-processing step by solving the full Hamiltonian in a Bloch-like basis constructed from Gaussian orbitals.  A key feature of the method is that it can incorporate combined translation--rotation, or screw, symmetries, enabling efficient simulations of chiral and helical systems with finite twist angles at essentially the same computational cost as systems described by pure translational symmetry. The approach is validated through calculations on linear and twisted $\mathrm{{-}C{\equiv}C{-}}$ and $\mathrm{{-}Li{-}F{-}}$ chains, as well as graphene, where total energies and band structures show excellent agreement with reference periodic calculations. To illustrate the ability of the method to treat three-dimensional systems, it is further applied to AA graphite, in which carbon atoms in adjacent graphene layers are aligned directly above one another, as well as helically stacked AA graphite structures. The Wannier-first framework provides a practical route for treating extended systems with nontrivial translational, rotational, and screw symmetries, and provides a natural foundation for the implementation of orbital-dependent functionals such as the Perdew--Zunger self-interaction correction.

\end{abstract}
\keywords{Wannier-first, Chiral systems, Density functional Theory, Self interaction corrections , AA Graphite}

\maketitle

\section{Introduction}\label{sec1} 
Chiral materials have attracted considerable attention because their handedness can influence electronic, magnetic, and spin-dependent phenomena.\cite{Bloom2024_ChemRev_CISS,Evers2022_AdvMater_CISS} A particularly important example is chiral-induced spin selectivity (CISS), in which electron transport through a chiral structure becomes spin selective even in the absence of conventional magnetic ordering.\cite{Bloom2024_ChemRev_CISS,Naaman2022_AnnuRevBiophys_51_99,Evers2022_AdvMater_CISS}. The CISS effect has stimulated growing interest in chiral molecular systems, helical polymers, covalent organic frameworks, molecular magnets, and biomolecular assemblies, where the interplay between structural chirality, electronic localization, and spin polarization can produce novel transport and magnetic behavior. \cite{Han2024_JACS_COF_CISS,Wang2025_NanoLett_ChiralMagnet,Getahun2023_APL_122_241903}


First-principles simulations of extended chiral systems are challenging because the underlying symmetry often involves not only translation, but also rotation. Examples include helical chains, screw-symmetric polymers, twisted layered materials, and chiral magnetic structures. Conventional periodic electronic-structure methods based solely on translational periodicity typically treat such geometries using large supercells that become computationally expensive or impractical for small twist angles, long helical pitches, or arbitrary rotational symmetries. As a result, there is strong motivation for electronic-structure methods that can treat screw symmetries directly within a minimal computational cell.

Most periodic electronic-structure methods are formulated in terms of canonical Bloch states and reciprocal-space representations.\cite{kohn1965,Kresse1996_CMS_6_15} Localized Wannier-like functions are then constructed afterward through unitary transformations of the Bloch eigenstates.\cite{wannier1937,marzari1997,Wannier90} These approaches are widely used and can provide highly accurate localized orbitals for periodic systems. However, obtaining well-localized Wannier-like functions generally requires dense $k$-point sampling together with additional localization procedures.\cite{Wannier90,Marzari2012_RMP_84_1419} More importantly, as noted above, extended systems possessing screw symmetry or arbitrary rotational periodicity can be difficult to represent efficiently within conventional methods based solely on translational periodicity because they may require very large supercells or may not admit a finite translational supercell at all.

Generalizations of Bloch's theorem have been developed to incorporate crystal
symmetries beyond simple translations and to treat nonstandard boundary conditions. Dobard\v{z}i\'c \textit{et al.} constructed symmetry-adapted Bloch
Hamiltonians using group-theoretical arguments,\cite{Dobardzic2015GeneralizedBloch} while Alase \textit{et al.} developed a generalized Bloch theorem for finite-range lattice Hamiltonians subject to arbitrary boundary conditions,\cite{Alase2017GeneralizedBloch} which was subsequently extended to higher-dimensional systems with surfaces and interfaces by Cobanera \textit{et al.}\cite{Cobanera2018GeneralizedBloch} These works illustrate
generalized Bloch constructions that incorporate additional crystal symmetries or nonstandard boundary conditions into lattice eigenvalue problems.

These generalized Bloch formulations and localized real-space approaches provide complementary ways of exploiting the structure of extended systems. The relative computational efficiency of real-space and reciprocal-space formulations depends on the system and on the quantities of interest. Calculations focused on total energies and charge densities, for example, have different computational requirements from those aimed at band structures, transport, or excitonic properties. Likewise, convergence with respect to the size of a localized real-space domain and convergence with respect to Brillouin-zone sampling are distinct numerical considerations associated with the chosen representation. The availability of complementary real- and reciprocal-space approaches can therefore be advantageous for different classes of problems. Within Bloch-based approaches, developments such as the optimized projection functions method have further automated the construction of maximally localized Wannier functions, including for bands with nontrivial topology.\cite{Mustafa2015OPFMWannier,Mustafa2016TopologicalWannier} Localized orthogonal orbitals are particularly useful for orbital-dependent methods, including self-interaction corrections (SIC),\cite{perdew1981,heaton_1987,mark_localization} and can also provide a useful representation for problems involving spatially separated electron and hole states, such as charge-transfer excitations.\cite{Baruah2009,Baruah2006}


In this work, we present a periodic DFT framework, referred to as the Wannier-first approach, in which Wannier-like orbitals are constructed directly from localized Gaussian basis functions without computing canonical Bloch-like states during the self-consistency cycle. The method builds on the "$\chi$-finding" formalism introduced by Pederson and Lin, who demonstrated the direct construction of localized Wannier-type orbitals for calculating the band structure of crystalline silicon.\cite{mark87} In the present work, this formalism is extended to one- and two-dimensional systems, as well as to systems where the fundamental symmetry operations involve translations plus rotations about the translation axis. The Wannier-like orbitals obtained are used to construct the self-consistent charge density, electrostatic potential, and per cell total energy for extended systems. Electronic band structures can be recovered in a subsequent post-processing step by solving the extended Hamiltonian in a basis of Bloch-like orbitals.

A key feature of the Wannier-first method is its natural incorporation of combined translation--rotation (screw) symmetries. This enables chiral and helical systems with arbitrary twist angles to be treated using the same real-space Wannier-domain construction employed for systems involving only translational symmetry, without requiring large supercells. The Wannier-like representation of the occupied space also provides a natural framework for orbital-dependent exchange--correlation functionals, including the Perdew--Zunger SIC.\cite{perdew1981}

In this paper, the accuracy and flexibility of the Wannier-first approach are demonstrated through applications to linear and twisted $\mathrm{{-}C{\equiv}C{-}}$ and $\mathrm{{-}Li{-}F{-}}$ chains, monolayer graphene, AA graphite and helically stacked AA (hAA) graphite. For the benchmark systems, the calculated total energies and band structures are in excellent agreement with reference periodic calculations.

In the next section, we outline the methodology.  That is followed by the computational details used in the test calculations.  We then present the results of the tests.  In the final section, we present our conclusions and an outlook for future work.

\section{Methodology}\label{sec2}

\begin{figure}
    \centering
    \includegraphics[width=0.6\linewidth]{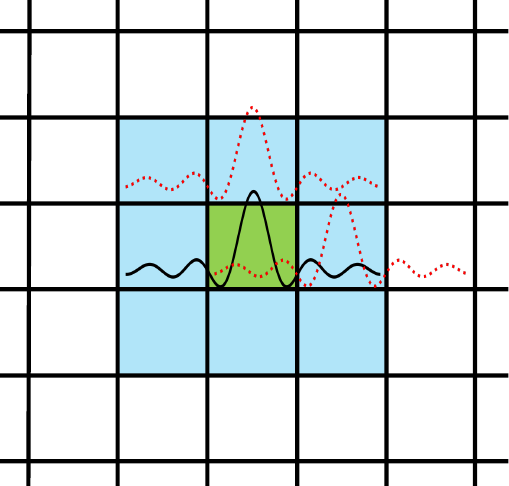}
\caption{Schematic representation of the Wannier domain (blue) surrounding the central site-zero cell (green). The solid black curve denotes a localized Wannier-like function associated with the central cell, while the dotted red curves represent orthogonal symmetry-related functions. Folding the density contributions of the central cell functions in neighboring cells in the Wannier domain back into the central cell recovers the total density of the extended system.}
    \label{wannierdomain}
\end{figure}

\begin{figure}
    \centering
    \includegraphics[width=0.8\linewidth]{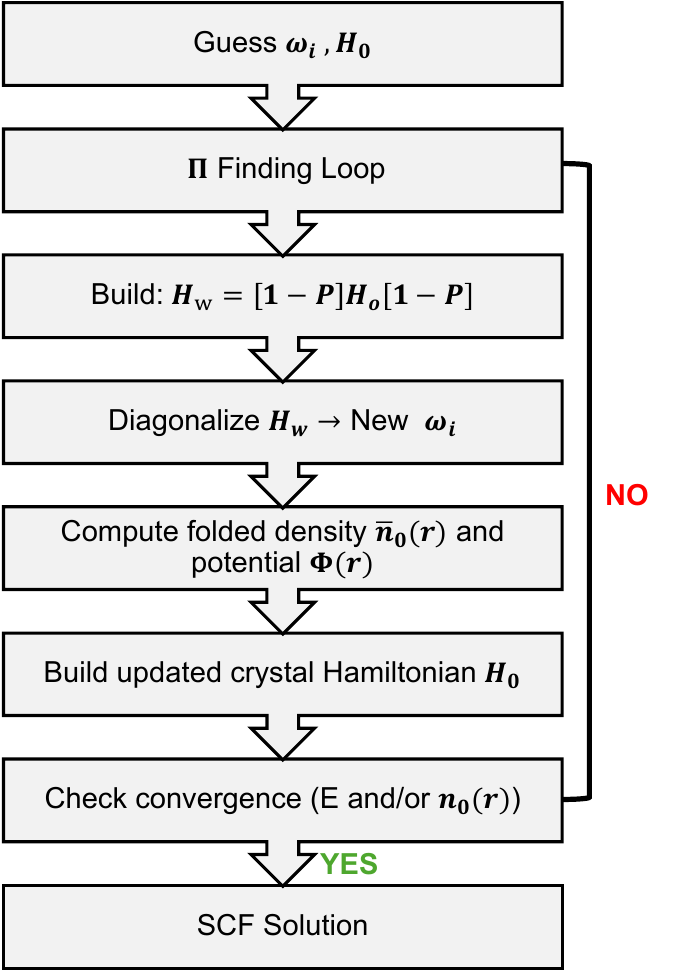}
    \caption{Flowchart of the self-consistent field (SCF) procedure used to obtain converged total energies and localized Wannier-like orbitals.}
    \label{scfcycle}
\end{figure}

\subsection{Wannier-First Framework}
The Wannier-first method builds on the $\chi$-finding formalism introduced by Pederson and Lin.~\cite{mark87}
We start from localized functions $\{\chi_i\}$ defined for a central unit cell ("site zero") where $i=1,\dots,L$ and $L$ is the number of  electrons associated with the central cell. These functions will be iteratively updated until they converge to the central cell Wannier-like functions $\omega_i$, but we follow Pederson and Lin~\cite{mark87} and refer to them as the $\chi$ functions in this section.  

We construct the $\chi$-finding operator ($\hat{\Pi}$): 

\begin{equation}{\label{eq1}}
\hat{\Pi} = \sum_{m\ne0}^{N-1}
\sum_{i=1}^{L} \left| \mathcal{T}_m \chi_i \right\rangle \left\langle \mathcal{T}_m \chi_i \right| - \tau \sum_{i=1}^{L} \left| \chi_i \right\rangle \left\langle \chi_i \right|.
\end{equation}
Here, $\mathcal{T}_m$ denotes a symmetry operation that may be either a pure lattice translation or a combined translation--rotation (screw) operation and $\tau$ is a positive parameter set equal to 1 in this work.  The $\hat{\Pi}$ operator is derived~\cite{mark87} from an error function that sums the squares of the overlaps of the central cell $\chi$ functions with their translations or translations+rotations. These overlaps are zero for Wannier-like functions. The purpose of the $\hat{\Pi}$ operator is to minimize the error function by updating the $\chi$ functions to make them orthogonal to their translated/translated and rotated counterparts.

The sum on $m$ in Eq.\ref{eq1} is over the unit cells included in the Wannier domain (WD), not including the central cell. The WD is the region of space where the amplitudes of the central cell Wannier-like functions are expected to be non-negligible. It consists of replicas of the central cell generated by the symmetry operations $\mathcal{T}_m$. Figure~\ref{wannierdomain} shows a schematic of the WD and the localized Wannier-like functions. In principle, the tails of a localized Wannier-like function extend over the entire crystal; however, their contribution to the charge density becomes negligible outside the WD.

The size of the central cell remains fixed, while convergence with respect to the real-space representation is controlled by systematically increasing the size of the WD. Increasing the WD incorporates contributions from progressively more distant symmetry-related cells until the calculated charge density and total energy converge to the desired accuracy. This real-space truncation convergence is distinct from conventional $\mathbf{k}$-point convergence, which controls Brillouin-zone sampling in reciprocal-space calculations.

Matrix elements of the $\hat{\Pi}$ operator are calculated in a basis that includes Gaussian orbitals centered on all of the atoms in the WD.  The matrix elements can be obtained analytically, using the properties of Gaussian functions.

The eigenvectors of the $\hat{\Pi}$ operator $\{|\chi_i\rangle\}$ can be classified into three groups according to their eigenvalues $\lambda_i$. The first group consists of $L$ eigenvectors with eigenvalues near $-\tau$. 
These are the updated central cell $\chi$ functions. A second group of $(N-1)L$ eigenvectors have positive eigenvalues and represent functions with non-zero overlap onto the  
translated/translated and rotated $\chi$ functions. The third group possesses zero eigenvalues and is orthogonal to both the site-zero and translated/translated+rotated $\chi$ functions. Note that because all three groups of functions are eigenfunctions of the same Hermitian operator, they are guaranteed to be mutually orthogonal.

The updated $\chi$ functions are orthogonal to the translated/ translated+rotated images of the $\chi$'s from the previous iteration, but they are not necessarily orthogonal to their own translations.  Thus, the construction of the $\hat{\Pi}$ operator is iterated until the updated $\chi$ functions are equal to the previous $\chi$s to some tolerance.  This is enforced in practice by monitoring the maximum value of $<\chi_i|\mathcal{T}_m\chi_i>$.
When this quantity drops below a prescribed threshold ($10^{-5}$ is used here), the $\chi$-finding loop is terminated. 

The states with positive eigenvalues are used to construct the projection operator $\hat{P} = \sum_{\lambda_i > 0}|\chi_i\rangle \langle \chi_i |$ and $\hat{P}$ is used in turn to construct the Wannier Hamiltonian
\begin{equation}\label{eq2}
\hat{H}_w = (1-\hat{P})\, \hat{H}_0 \, (1-\hat{P}).
\end{equation}
\noindent Here $\hat{H}_0$ is the full Hamiltonian for the periodic system. The operator $(1-\hat{P})$ projects $\hat{H}_0$ onto the subspace of the space spanned by the WD basis functions that is orthogonal to the translated/translated+rotated Wannier-like functions. It can be shown \cite{mark87} that (at convergence) the site-zero Wannier-like functions are the lowest eigenfunctions of $\hat{H}_w$. 
Diagonalization of $\hat{H}_w$ therefore yields updated approximations of the localized Wannier-like functions that are subsequently used to update the  charge density and $\hat{H}_0$. 

The entire procedure is repeated as illustrated in Fig.~\ref{scfcycle} until the total energy converges below a chosen tolerance $\epsilon$. In the present work, $\epsilon$ is set to $10^{-5}$ Hartree. We find that a typical calculation converges in about 10 cycles. 
We stress that, unlike conventional Wannier approaches, the present formulation constructs localized Wannier-like functions directly, without first computing Bloch-like states or performing reciprocal-space operations during the SCF cycle.

\begin{table*}[]
\centering
\caption{Band gaps $E_g$ and total-energy differences $\delta E$ for linear, chiral, graphene, and layered graphene systems. For $\mathrm{{-}C{\equiv}C{-}}$, $\mathrm{{-}Li{-}F{-}}$, and graphene, $\delta E$ compares the Wannier-first results with reference periodic calculations performed using the Gaussian software\cite{g16} and the same basis functions as in the Wannier-first calculations. The $\delta E$ values for helically stacked graphite (hAA graphite) represent the total energy differences relative to AA graphite in the Wannier-first method.}
\begin{tabular}{l c c c  c}
\toprule
\multicolumn{2}{c}{System} & \multicolumn{2}{c}{ $E_g$ (eV)} & $\delta E$ \\
\cmidrule(lr){1-2} \cmidrule(lr){3-4} \cmidrule(lr){1-2} \\
System & Twist Angle & Wannier-First &  Gaussian &  (eV)\\
\hline
$\mathrm{{-}C{\equiv}C{-}}$  & 0$^\circ$ & 2.96   & 2.96    &  $1.88\times10^{-4}$\\
               & 60$^\circ$ & 3.23  & 3.23    & $3.23\times10^{-4}$\\
$\mathrm{{-}Li{-}F{-}}$ & 0$^\circ$ & 3.86  & 3.86& $3.79\times10^{-4}$\\
              & 60$^\circ$ & 3.62 & 3.62 &$7.04\times10^{-4}$\\
Graphene      & 0$^\circ$ & 0.0  & 0.0 &$6.82\times10^{-3}$\\  
hAA Graphite &  0$^\circ$ & 0.0  &$-$&0.0\\
              &  5$^\circ$ & 0.0  &$-$&${2.30\times10 ^{-2}}$\\
              & 15$^\circ$ & 0.0  &$-$&${2.36\times10 ^{-2}}$\\
              & 25$^\circ$ & 0.0  &$-$&${3.92\times10 ^{-2}}$\\           
\hline
\end{tabular}
\label{tab2}
\end{table*}

\subsection{Total Energy}
Within the Kohn--Sham formalism, the total energy is written as
\begin{equation}\label{eq3}
E = T_s + E_H + E_{eN} + E_{xc} + E_{NN},
\end{equation}
where $T_s$ is the kinetic energy of the non-interacting electrons, $E_H$ is the Hartree energy, $E_{eN}$ is the electron--ion interaction energy, $E_{xc}$ is the exchange--correlation energy, and $E_{NN}$ is the ion--ion interaction energy. In this work, $E_{xc}$ is evaluated within the local density approximation (LDA).\cite{Perdew1992}

Once the localized Wannier-like functions associated with the central cell (site-zero), $\omega_{i}$ ($i = 1,L$) are known, the total charge density can be evaluated.  In the following $ n(\mathbf{r}) = \sum_{a}n_{0}({\mathbf{r}-\mathbf{T}_a})$ where $n_{0}(\mathbf{r})$ is the density corresponding to the site-zero Wannier-like functions and $\mathbf{T}_{a}$ is a lattice translation vector. This expression for the total density accounts for the fact that $n_0(\mathbf r)$ extends beyond the central unit cell into neighboring cells. By symmetry, the Wannier-like functions in neighboring cells are obtained from the site-zero functions by pure translations or, for screw-symmetric systems, by combined translation--rotation operations.  This means that the contributions of Wannier-like functions associated with neighboring cells to the central cell density can be obtained by folding the density associated with the central cell Wannier functions in the neighboring cell back to symmetry-equivalent positions in the central cell. The central cell Wannier-like functions contain all the essential information.

The total energy per cell of a periodic system can be expressed as 
\begin{equation}\label{eq3}
    \begin{split}
\frac{E}{N} =
&\sum_{i=1}^{L} 
\left\langle \omega_{i} \left| -\frac{\nabla^2}{2} \right| \omega_{i} \right\rangle
 - \sum_{j,a \in {WD}}\int n_0(\mathbf{r}) 
\frac{Z_j}{|\mathbf{r} - (\mathbf{T}_a + \mathbf{R}_j)|} d\mathbf{r}
\\
&+ \frac{1}{2} \sum_{a \in {WD}} \iint \frac {n_{0}(\mathbf{r})n_{a}(\mathbf{r}')}{|\mathbf{r}-\mathbf{r}'|}d\mathbf{r}' d\mathbf{r} +\int_{\Omega_{0}}n(\mathbf{r})\varepsilon_{xc}[n(\mathbf{r}),0]d\mathbf{r} \\
&+ \frac{1}{2} \sum_{i,j,a \in {WD}}
\frac{Z_i Z_j}
{|\mathbf{R}_i - (\mathbf{T}_a + \mathbf{R}_j)|} \\
&- \frac{1}{2} \sum_{i,a \notin {WD}} Z_{i}\Phi_{0}(\mathbf{R}_i + \mathbf{T}_a)  + \frac{1}{2} \int n_{0}(\mathbf{r})\sum_{a \notin {WD}} \Phi_{a}(\mathbf{r}) d\mathbf{r} 
\end{split}
\end{equation}
In the term corresponding to the exchange-correlation part of the energy above the integration is confined to the central cell volume $\Omega_0$. In the last two terms of this expression, we introduce $\Phi$, the Coulomb potential generated by all charges (electrons + protons) associated with a given cell. For the central cell, this potential is defined as
\begin{equation}
\Phi_{0}(\mathbf r)
= \int
\frac{n_{0}(\mathbf r')}
{|\mathbf r-\mathbf r'|}
 d\mathbf r'
-
\sum_{j}\frac{Z_j}
{|\mathbf r-\mathbf{R}_j|}
\end{equation}
where the sum over $j$ runs over all nuclear sites in the central cell. For points outside of the WD, $\Phi_{0}$ can be approximated by a multipole expansion, 
\begin{equation}
    \Phi_{0}(\mathbf{r})=\sum_{l,m}\frac{S_{lm}Y_{l,m}(\theta,\phi)}{r^{l+1}}.
\end{equation}
The multipole coefficients $S_{lm}$ are obtained by evaluating $\Phi_{0}$ exactly on a sphere of radius $R$ surrounding, but well outside the boundaries of the central cell. In this work, we use $R = 60$ Bohr.  Using the orthonormality of the spherical harmonics ($Y_{lm}$'s),
\begin{equation}
    S_{lm} = R^{l+1}\iint\Phi_{0}(R,\theta,\phi)Y_{lm}^{*}(\theta,\phi) \text{sin}(\theta) d\theta d\phi.
    \label{eq7}
\end{equation}
For the systems considered here, we find that including terms up to $l=6$ is sufficient to achieve good accuracy.  Since the central cell is charge neutral, the monopole term $S_{00}$ is constrained to be exactly zero.

For systems with screw symmetry, the electrostatic contribution from cells outside the WD must also incorporate the corresponding rotational symmetry. In this case, the multipole moments associated with each symmetry-related cell are rotated relative to those of the central cell by an angle determined by the screw operation connecting the two cells. To evaluate the rotated multipole coefficients, the spherical harmonics $Y_{lm}(\theta,\phi)$ are first evaluated on an angular grid rotated by the same angle as the corresponding cell. Eq.~\ref{eq7} is then used to construct the rotated set of multipole coefficients $S_{lm}$ for each symmetry-related image. These rotated multipole moments are subsequently used to evaluate the long-range electrostatic potential arising from cells outside the finite WD while preserving the underlying screw symmetry of the system.
\subsection{Band Structure}

Following convergence of the SCF cycle, the converged potential of the extended system is used to evaluate the electronic band structure in a separate post-processing step. In the present approach, canonical Bloch-like states are not employed during the SCF procedure. Instead, Bloch basis functions are constructed from the localized Gaussian basis orbitals $f_n(\mathbf r)$ associated with the central cell:

\begin{equation}
\phi_{n\mathbf{k}}(\mathbf r)
=
\frac{1}{\sqrt{N}}
\sum_m
e^{i\mathbf{k}\cdot\mathbf{T}_m}
\, f_n\!\left(\mathcal{T}_m^{-1}\mathbf r\right),
\end{equation}

where $N$ is the number of symmetry-related cells included in the summation. It can be shown that $\phi_{n,\mathbf{k}}$ satisfies a generalized Bloch condition and that the Hamiltonian and overlap matrices in this basis are block diagonal in {$\mathbf{k}$} and have the form: 

\begin{equation}
\langle \phi_{n\mathbf k}| \hat{H}|\phi_{n'\mathbf k}  \rangle  
= \sum_m e^{i\mathbf{k}\cdot\mathbf{T}_m} H_{nn'}(\mathcal{T}_m),
\end{equation}

\begin{equation}
\langle \phi_{n\mathbf k}| \phi_{n'\mathbf k}  \rangle  
= \sum_m e^{i\mathbf{k}\cdot\mathbf{T}_m} S_{nn'}(\mathcal{T}_m).
\end{equation}

Here,
\begin{equation}
H_{nn'}(\mathcal{T}_m) = \left\langle f_n \middle| \hat H \middle| \mathcal{T}_m f_{n'} \right\rangle,
\end{equation}

\begin{equation}
S_{nn'}(\mathcal{T}_m) = \left\langle f_n \middle| \mathcal{T}_m f_{n'} \right\rangle,
\end{equation}
represent the Hamiltonian and overlap matrix elements between the central cell and its symmetry-related neighboring cells in the localized basis representation. In principle, these matrix elements extend over all symmetry-related images of the system. In practice, however, their magnitudes decay rapidly with distance due to the localized nature of $f_{n}(\mathbf{r})$. Consequently, the matrix elements are evaluated only within the finite Wannier domain (WD). 

Importantly, the band-structure calculation does not require repeating the SCF procedure for each $\mathbf{k}$ point. Once the self-consistent potential has converged, the full sets of $H_{nn'}(\mathcal{T}_m)$ and $S_{nn'}(\mathcal{T}_m)$ are evaluated within the finite WD and reused for all $\mathbf{k}$ points. The full $H(\mathbf{k})$ and $S(\mathbf{k})$ matrices, including both diagonal and off-diagonal elements in the localized basis, are then obtained through the Fourier sums above. Consequently, evaluating additional $\mathbf{k}$ points requires only the construction and diagonalization of these finite matrices and does not involve additional self-consistent electronic-structure calculations. 

The electronic band energies $\epsilon_n(\mathbf k)$ are obtained by solving the generalized eigenvalue problem
\begin{equation}
H(\mathbf k)\, \mathbf C_{n\mathbf k} =
\epsilon_n(\mathbf k)\, S(\mathbf k)\, \mathbf C_{n\mathbf k}.
\end{equation}

\section{Computational Details}{\label{sec:computational_details}}
We implemented the Wannier-first method in the Naval Research Laboratory Molecular Orbital Library (NRLMOL) code.\cite{nrlmol_pederson90,nrlmol_jack90,nrlmol_porezag99,nrlmol_pederson2000} All calculations reported here were performed using minimal basis sets taken from the default NRLMOL Gaussian basis sets.\cite{nrlmol_porezag99} For carbon and lithium atoms,  two $s$-type and one $p$-type orbital were employed, based on 12 and 10 single Gaussian orbitals, respectively, while for fluorine atoms three $s$-type and two $p$-type functions were used, based on 14 single Gaussians. These small basis sets proved adequate for testing purposes. 

The Wannier-first procedure requires an initial set of approximate localized Wannier-like functions associated with the central unit cell. In this work, these trial orbitals were generated by solving for the eigenfunction of an {\em ad hoc} Hamiltonian consisting of overlapping atomic potentials for the central cell atoms. 

In some cases, a small, spin-symmetry-breaking bias was added to the starting Hamiltonian in order to obtain localized orbitals with the expected character of bond-centered valence Wannier-like functions. For example, in the linear carbon chain, the central cell contains six electrons of each spin. For a given spin, two electrons can be associated with $1s$ core states on the two carbon atoms, while three electrons participate in the central C$\equiv$C triple bond. The remaining spin-up electron must be assigned to one of the two bond centers that connect the central cell to a neighboring cell. We therefore introduce a small bias in the ad hoc starting Hamiltonian  to better localize a spin-up eigen state on one of these bonds and the spin-down eigen state on the other. This symmetry breaking is used only to construct the initial localized trial functions. During the SCF procedure, the density and orbitals are relaxed variationally, and the final converged solution restores the physical spin symmetry of the system. The disappearance of the initial spin imbalance during the SCF cycle is documented in Table S1 of the {\textbf{Supporting Information}.} 

The WD size used in each calculation was increased systematically until the integrated site-zero charge and total energy were converged. Table S2 of the Supporting Information illustrates this convergence for the $\mathrm{{-}C{\equiv}C{-}}$ and $\mathrm{{-}Li{-}F{-}}$ chains.

The structural parameters defining the central cell for the one-dimensional 
test systems studied for this work are summarized in Table S3 of {\bf Supporting Information}. For the carbon chain, the central-cell $\mathrm{{-}C{\equiv}C{-}}$ bond length was fixed at 1.22~\AA~  corresponding to a triple bond between these atoms, with a lattice translation of 2.75~\AA~ that corresponds to single bonds between the central cell atoms and atoms in adjacent cells. For the $\mathrm{{-}Li{-}F{-}}$ chain, a bond length of 1.63~\AA~  and a translation of 3.26~\AA~ were used. This corresponds to bonds of the same length between each Li and F.  

For the linear chains, all atoms lie on the $x$-axis.  To create helical chains, we move the central cell atoms off the $x$-axis in the $y$ direction.  Atoms in other cells are generated by using a symmetry operation that combines a translation in $x$ plus a fixed rotation about the $x$-axis.  The rotation does not change bond distances between atoms within a given cell, but stretches the bonds connecting atoms in neighboring cells.

For graphene, a two-atom central cell was used, with the atoms separated by a distance of 1.42~\AA. The coordinates of these atoms and the periodic translation vectors for graphene are given in Table S3 of the {\bf Supporting Information}. Using this structure for graphene, we created AA graphite by adding a third translation vector in the out-of-plane direction. In AA stacking, each C atom of a translated graphene plane lies directly above the corresponding atom in the central plane.  We optimized the inter-layer separation via total-energy minimization within the Wannier-first framework. With the minimal basis used in these calculations, the lowest-energy stacking distance was found to be 2.80~\AA.  

The hAA graphite structures were generated by combining the out-of-plane translation with a fixed rotation angle between successive layers. The rotation was performed about an axis passing through the origin of the central cell. This construction produces an infinite helical stack in which each layer is related to the preceding layer by the same translation+rotation operation. Representative twist angles of $5^\circ$, $15^\circ$, and $25^\circ$ were considered. 

Electronic band structures for the test systems were evaluated along selected high-symmetry paths in reciprocal space. For the linear and helical chains, this is simply the path from $\Gamma$ to $X$.  For graphene and the related graphite systems, the $\Gamma$--K--M--$\Gamma$ path was used to describe the in-plane dispersion. Details of the k-point path used for graphene and AA graphite are provided in Table S4 of the {\bf Supporting Information}.

To test the Wannier-first method, we compare its results to values from reference calculations.  First, for the linear and helical chains, we obtained total energies and electronic densities of states from  systematic cluster calculations containing increasing numbers of repeat units, in all cases capping the ends of the chain with H atoms to preserve the correct single bond character at the ends. These calculations are carried out with the NRLMOL code, using the same basis as assigned to the central cell in the Wannier-first calculations. To predict the total energy per unit cell, we subtract the total energy of a chain with $N-2$ units from that with $N$ units and divide the result by two to get the energy per unit.  Subtracting energies in this way removes the contribution of the chain ends from the total energy. Fig. ~S1(a) of the Supporting Information shows that the cluster-based total energy per unit for the carbon chain converges smoothly with increasing $N$, reaching a limiting value by $N = 24$. Fig.~S1(b) further shows that the relative total energies obtained with the Wannier-first method agree closely with the cluster-based approximation over the full range of twist angles considered.

A second reference calculation made use of the Gaussian software package \cite{g16} with periodic boundary conditions. These calculations solve for the electronic structure using canonical Bloch functions and k-space methods. The advantage of using this particular reference is that the same Gaussian basis functions used for the central cell in the Wannier-first calculations are used in the corresponding reference calculations, minimizing differences associated with the choice of basis. 
Total energies, band structures, and densities of states obtained from the two approaches are then compared directly.

Finally, for AA and hAA graphite structures, additional reference calculations were performed using plane-wave DFT as implemented in Quantum ESPRESSO code.\cite{Giannozzi2009QE,Giannozzi2017QE} A kinetic-energy cutoff of 50 Ry was used for the plane-waves.

\section{Results and Discussion}\label{sec3}

\subsection{ Linear and helical chains}

\begin{figure*}
    \centering
    \includegraphics[width=1.0 \linewidth]{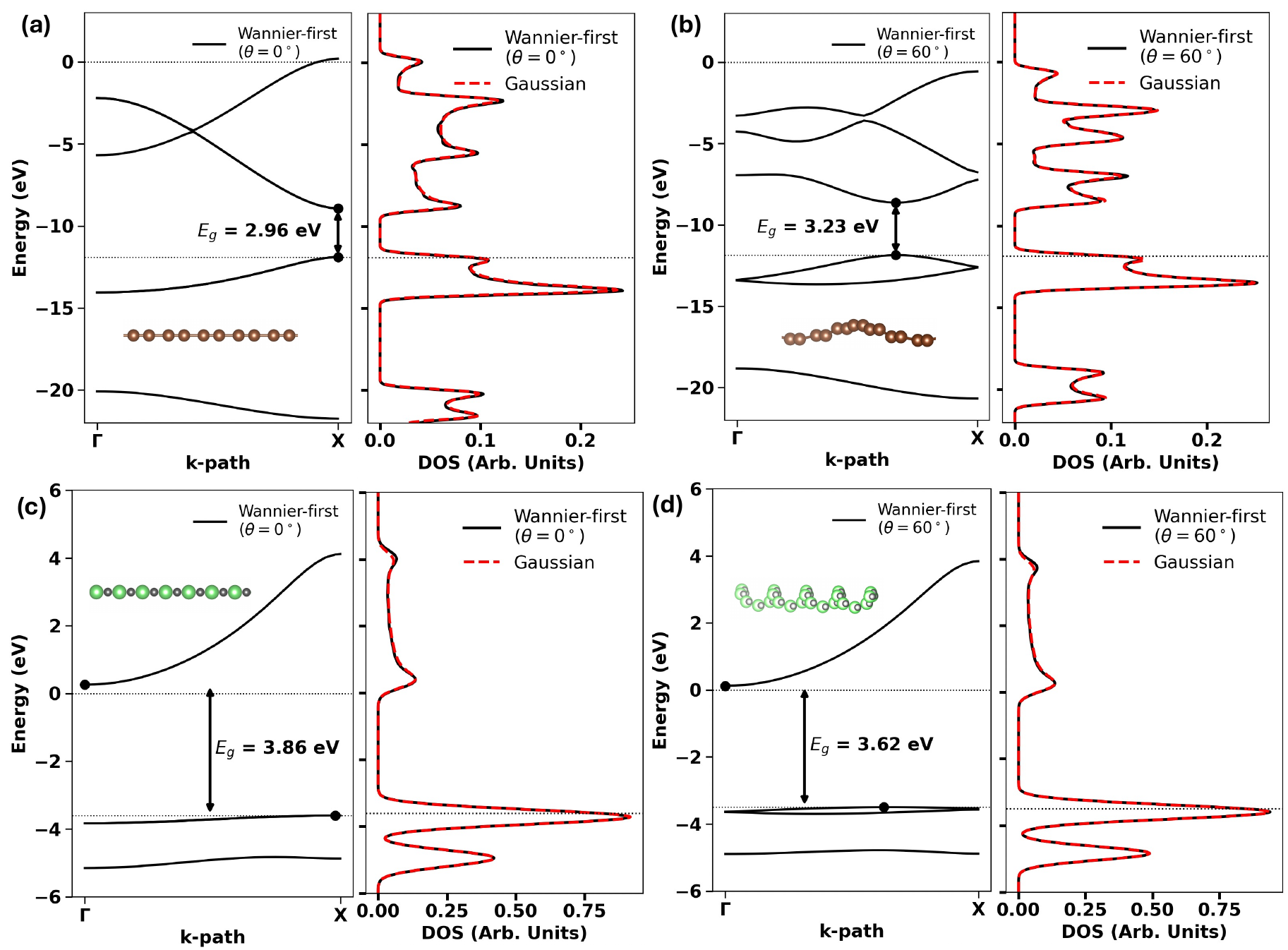}
    \caption{Band structures and densities of states (DOS) obtained using the Wannier-first method and a periodic version of the Gaussian code\cite{g16} for (a) linear$\mathrm{{-}C{\equiv}C{-}}$ chain, (b) twisted $\mathrm{{-}C{\equiv}C{-}}$ chain, (c) linear $\mathrm{{-}Li{-}F{-}}$ chain, and (d) twisted $\mathrm{{-}Li{-}F{-}}$ chain. The twist angle ($\theta$) is $60^\circ$.}
    \label{ccband}
\end{figure*}

\begin{figure}
    \centering
    \includegraphics[angle=0,width=\linewidth]{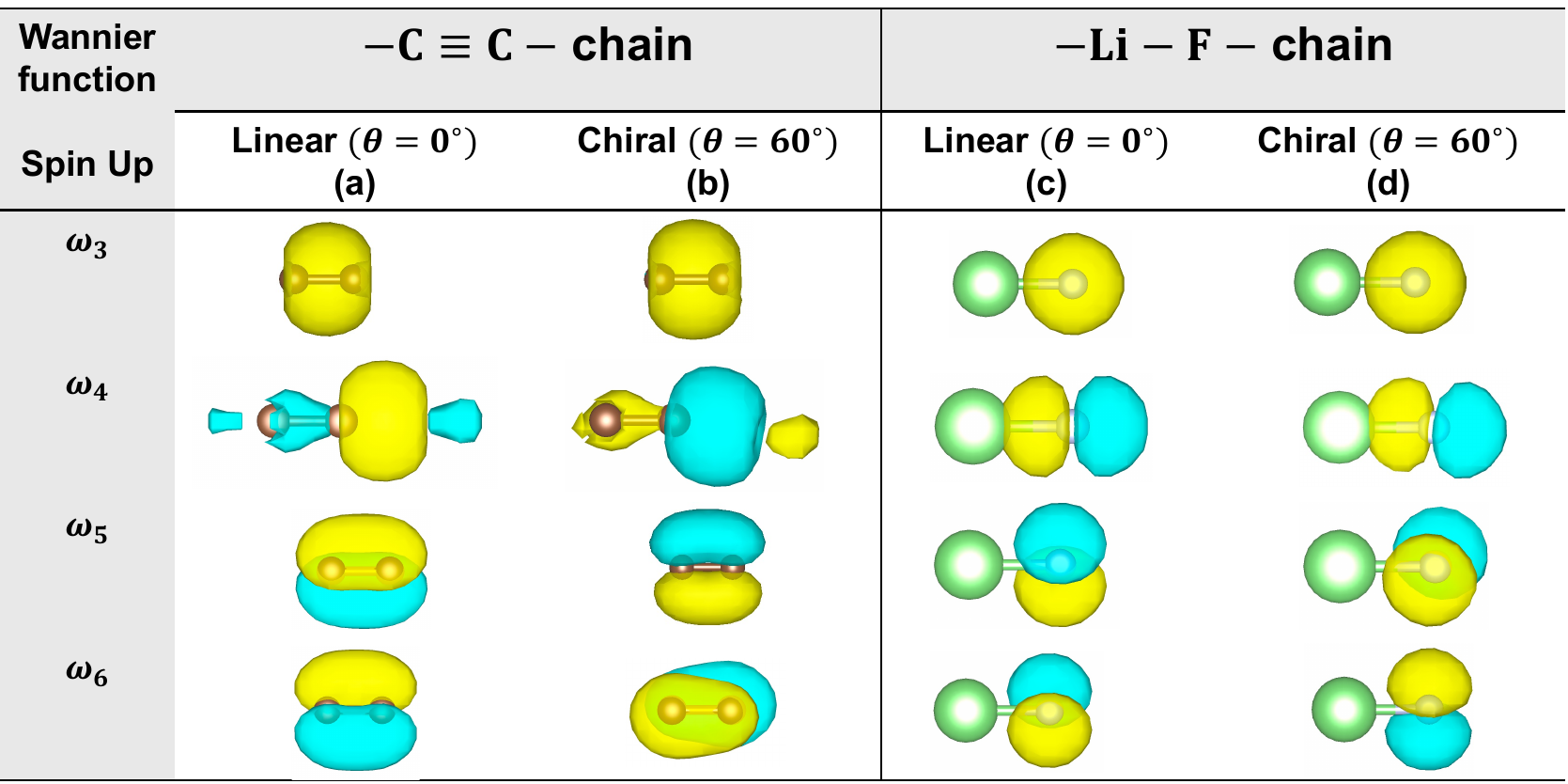}
     \caption{Spin-up Wannier-like functions $\omega_3$--$\omega_6$ for (a) linear and (b) chiral ($\theta = 60^\circ$) $\mathrm{{-}C{\equiv}C{-}}$ chains, and (c) linear and (d) chiral ($\theta = 60^\circ$) $\mathrm{{-}Li{-}F{-}}$ chains. Isosurfaces are plotted at a value of $0.01~\text{\AA}^{-3}$. Yellow and blue surfaces denote opposite phases of the Wannier-like functions. Carbon, lithium, and fluorine atoms are shown in brown, light green, and pale yellow, respectively.}
    \label{linearwf}
\end{figure}

We first demonstrate the accuracy of the Wannier-first approach for one-dimensional systems by considering linear and chiral $\mathrm{{-}C{\equiv}C{-}}$ and $\mathrm{{-}Li{-}F{-}}$ chains. These are useful test systems because they feature diverse bonding types and symmetry-related characteristics. Figure~\ref{ccband} shows the band structures and corresponding density of states (DOS) for both linear ($\theta = 0^\circ$) and helical ($\theta = 60^\circ$) configurations. As shown in panels (a)--(d), the Wannier-first DOS are indistinguishable from the Gaussian-based\cite{g16} periodic calculations over the entire energy range. 

Representative spin-up Wannier-like functions $\omega_3$--$\omega_6$ for the linear and helical chains are shown in Fig.~\ref{linearwf}. For the $\mathrm{{-}C{\equiv}C{-}}$ chain, the Wannier-like functions display the expected covalent bonding character. The orbitals can be classified according to their symmetry with respect to the chain axis: $\sigma$-like orbitals are symmetric under rotation about the axis, whereas $\pi$-like orbitals have lobes that are perpendicular to the axis and a node at the axis. This classification is clearly visible in the linear chain, where the localized orbitals retain the symmetry expected for a carbon--carbon triple bond. For the chiral carbon chain, the corresponding Wannier-like functions remain localized but rotate from one unit cell to the next, consistent with the imposed screw symmetry.

For the $\mathrm{{-}Li{-}F{-}}$ chain, the Wannier-like functions exhibit ionic character, with localized charge concentrated primarily around the fluorine atom and weaker amplitude near lithium. The linear and chiral configurations show similar localization patterns, while the chiral case again follows the rotation between neighboring cells. 

A quantitative comparison of total energies and band gaps is given in Table~\ref{tab2}, where the band gaps obtained using the Wannier-first method can be seen to be in excellent agreement with the corresponding values from periodic calculations performed using the Gaussian software.\cite{g16} The deviation in total energy per unit cell is on the order of $10^{-4}$~eV, demonstrating that the Wannier-first framework accurately reproduces both the orbital-energy spectrum and the total energy of the periodic system. For the carbon chain, the linear structure consists of alternating single and triple bonds and exhibits a direct band gap of 2.96~eV. Introducing a $60^\circ$ rotational component modifies the inter-cell geometry and increases the nearest-neighbor separation, resulting in a larger direct band gap of 3.23~eV. This increase reflects reduced overlap between neighboring $\pi$ orbitals. In addition, the fifth and sixth bands, which are degenerate by symmetry in the linear configuration, become split in the chiral structure due to 
the reduced symmetry of the helical chain.

The electronic structure of the $\mathrm{{-}Li{-}F{-}}$ chain reflects the predominantly ionic character of the bonding. The occupied valence bands are largely derived from localized fluorine $2p$ orbitals, whereas the low-lying conduction bands have greater lithium character and are more spatially diffuse. In the linear chain, symmetry preserves degeneracies among fluorine-centered states with equivalent orientations relative to the chain axis. Introducing screw symmetry modifies these symmetry relationships by rotating neighboring unit cells relative to one another. As a result, the overlap between fluorine-centered orbitals becomes anisotropic, leading to a small splitting and upward shift of the valence bands, while the conduction bands remain comparatively unchanged. This selective modification of the valence manifold produces the observed reduction in the indirect band gap. 
 
It is worth noting that the periodic reference calculations for the chiral chains require the use of supercells that reflect the periodicity of the helical systems.  
For the $\theta = 60^\circ$ structures, 
a sixfold supercell is required. 
In contrast, the Wannier-first approach 
uses a WD that has the same number of cells as in the pure translation case.  Furthermore, the Wannier-first method can be used to treat helical chains with small or irrational rotation angles for which the corresponding supercell would be very large or non-existent. 

\subsection{Graphene: Two-Dimensional Periodic Benchmark}

\begin{figure*}
    \centering
    \includegraphics[width=0.8\linewidth]{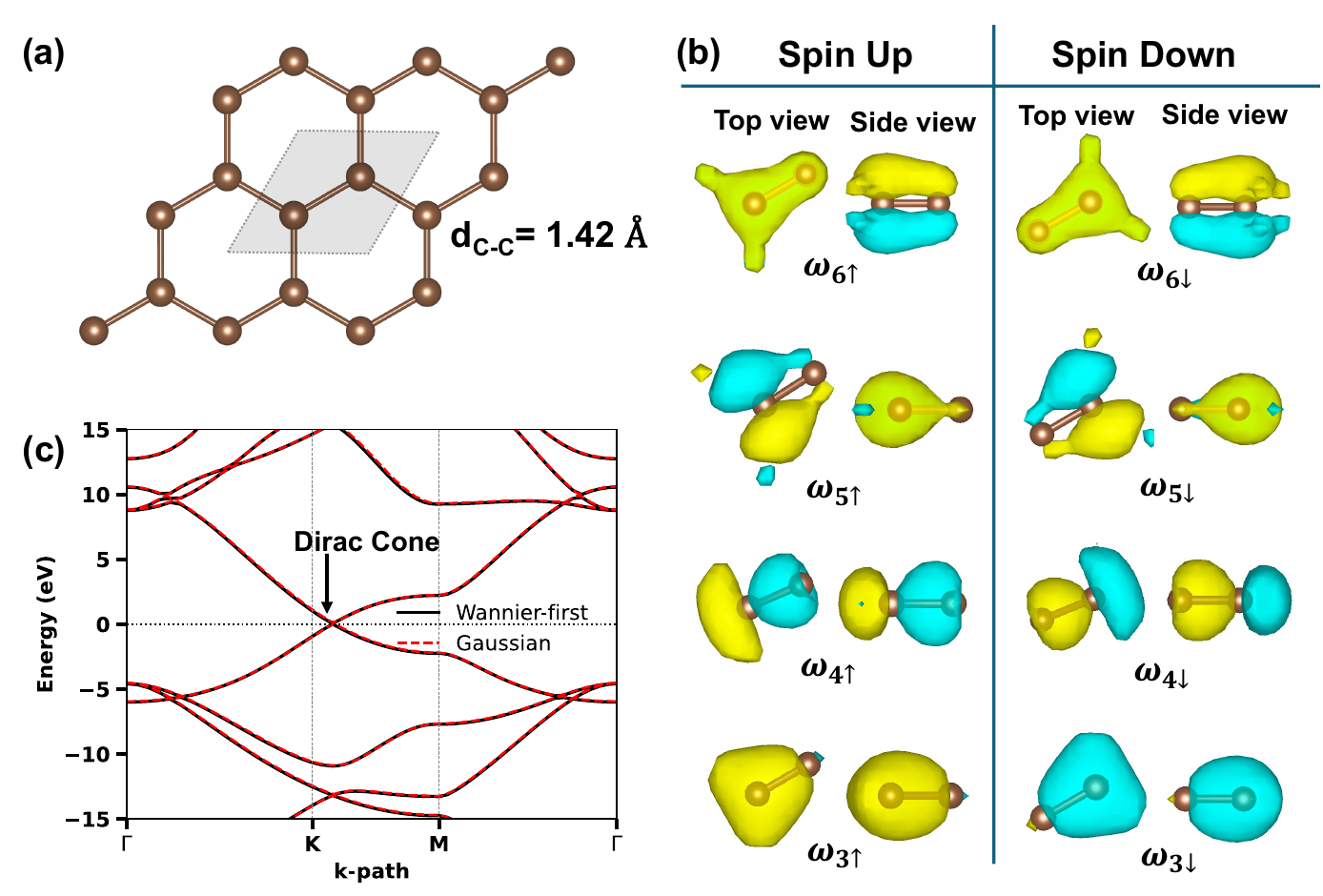}
    \caption{(a) Central cell with two carbon atoms (C–C distance 1.42 Å). (b) Spin-resolved Wannier-like functions (top and side views) corresponding to C 2s and 2p orbitals. (c) Band structure comparison between the Wannier-first approach and a Gaussian-based periodic calculation. Isosurfaces are plotted at a value of $0.01~\text{\AA}^{-3}$. Yellow and blue surfaces denote opposite phases of the Wannier-like functions.}
    \label{graphene_band}
\end{figure*}

We next validate the Wannier-first approach for a two-dimensional system by considering graphene, which provides an ideal test case because of its well-established electronic structure with a symmetry-driven linear band crossing at the K point. This Dirac cone is sensitive to the equivalence of the two carbon atoms in the unit cell and can be artificially gapped if the underlying lattice symmetry is not preserved. \cite{CastroNeto2009_RMP_Graphene,Wallace1947_PR_Graphite}

The graphene structure is shown in Fig.~\ref{graphene_band}(a) and Wannier-like functions for graphene are shown in Fig.~\ref{graphene_band}(b), plotted at an isosurface value of $0.01~\text{\AA}^{-3}$. The orbitals $\omega_3$--$\omega_6$ display the expected character of the $sp^2$ bonding network and the out-of-plane $p_z$ orbitals that form the $\pi$ and $\pi^\ast$ bands. The top and side views show that the Wannier-first procedure produces well-localized orbitals with clear chemical character while preserving the symmetry of the honeycomb lattice.

The equivalence of the spin-up and spin-down charge densities is consistent with the non-magnetic ground state of graphene. Although a small spin-symmetry-breaking bias is introduced in the initial trial Wannier-like functions associated with the central cell, the SCF procedure restores the overall spin symmetry. The final converged solution has zero net magnetization, with equivalent spin-up and spin-down densities.

The electronic band structure obtained from the Wannier-first approach is compared with the reference Gaussian-based periodic calculation in Fig.~\ref{graphene_band}(c).\cite{g16} The two methods are in excellent agreement throughout the Brillouin zone. In particular, the characteristic Dirac cone at the K point is reproduced accurately by the Wannier-first method, with the linear dispersion and band crossing preserved without any artificial gap opening.

Although the band structures are nearly identical, the total energy difference between the Wannier-first and Gaussian \cite{g16} calculations is slightly larger than for the one-dimensional systems, amounting to approximately $10^{-4}$ Hartree per unit cell. This small discrepancy originates in part from the finite size of the WD used to represent the localized orbitals and electrostatic interactions.   
To quantify the effect of domain size, we systematically increased the number of neighboring shells included around the central cell. For a two-dimensional system, a domain extending from $-n$ to $n$ along both lattice vectors contains $(2n+1)^2$ unit cells. As the domain size is increased from two to six shells, corresponding to 25 to 169 cells, the total energy and the site-zero Wannier charge converge systematically toward their asymptotic values (Table~\ref{tab3}).  In particular, the integrated central cell charge approaches the correct value of 12 electrons, and the residual gap at the Dirac point decreases steadily. We note that the remaining differences between the Wannier-first and Gaussian reference total energies may be due to numerical differences between the codes, for example, the use of different integration grids for computing total energies.

\subsection{AA and Helically Stacked AA Graphite }
\begin{figure}
    \centering
    \includegraphics[width=0.95 \linewidth]{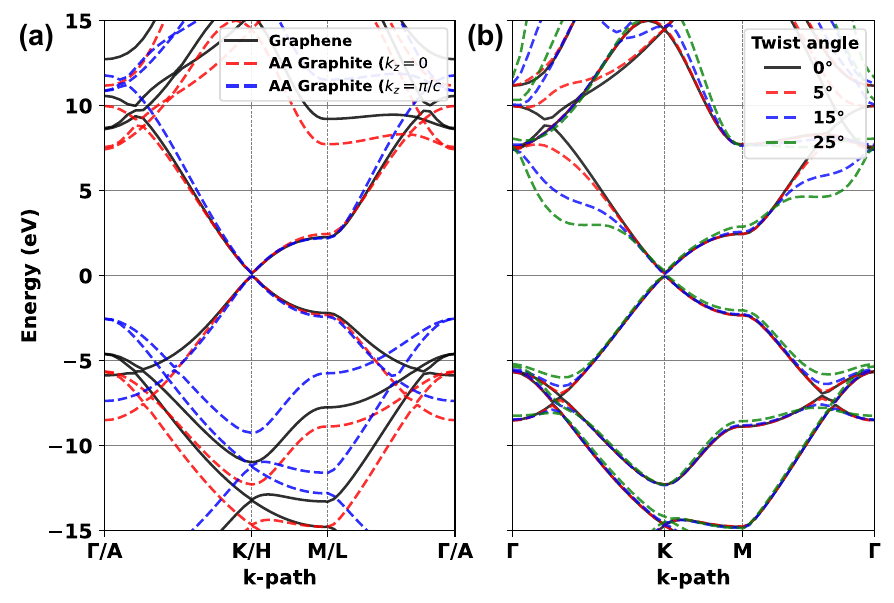}
    \caption{(a) Comparison of the electronic band structures of monolayer graphene and AA graphite. The black solid lines correspond to monolayer graphene, while the red and blue dashed lines represent AA graphite evaluated at $k_z=0$ along the in-plane path $\Gamma$--K--M--$\Gamma$ and at $k_z=\pi/c$ along the symmetry-equivalent path A--H--L--A, respectively. 
Inter-layer $\pi$--$\pi$ coupling introduces bonding and antibonding splittings and a finite dispersion along the stacking direction while preserving the Dirac crossing at K. (b) Band structures of helically stacked AA (hAA) graphite for twist angles of $0^\circ$, $5^\circ$, $15^\circ$, and $25^\circ$ between adjacent layers. The $0^\circ$ case corresponds to AA graphite. Only bands corresponding to $k_z=0$ are shown.  The Dirac crossing remains preserved for all twist angles, while the higher-energy bands exhibit progressive changes due to the rotational modulation of inter-layer coupling. In all panels, energies are referenced to the valence-band maximum (VBM), which is set to zero.}
    \label{fig_twisted_band}
\end{figure}

\begin{table}[t]
\centering
\caption{Convergence of the total-energy difference $\delta E$, site-zero Wannier charge, and residual Dirac-point gap with respect to Wannier domain size for graphene. Here, $\delta E$ is the difference between the Wannier-first total energy for a given shell size and the reference energy obtained from the Gaussian-based periodic calculation. The Wannier domain is defined by including $n$ shells of neighboring cells around the central cell.}
\label{tab3}
\begin{tabular}{c c c c}
\toprule
Shells ($n$) & $\delta$E (eV) & Total Charge (e) & Gap at K (eV) \\
\hline
2 & 9.02 $\times10^{-1}$ & 11.98046 & 0.32 \\
3 & 2.32 $\times10^{-1}$ & 11.99563 & 0.15 \\
4 & 4.01 $\times10^{-2}$ & 11.99939 & 0.07 \\
5 & 1.45 $\times10^{-2}$ & 11.99982 & 0.06 \\
6 & 6.88 $\times10^{-3}$ & 11.99993 & 0.05 \\
\hline 
\end{tabular}
\end{table}

We next investigate inter-layer coupling by extending graphene to a periodically stacked structure along the out-of-plane direction. For this test, the two-carbon central cell is translated along $z$, producing AA graphite, in which each carbon atom lies directly above its counterpart in adjacent layers. Figure~\ref{fig_twisted_band}(a) compares monolayer graphene with AA graphite evaluated at $k_z=0$ and $k_z=\pi/c$, where $c=2.80$~\AA\ is the optimized inter-layer separation obtained from Wannier-first total-energy minimization as shown in Fig. S2 of the Supporting Information.  The k-point paths used for graphite in this figure represent equivalent paths on different planes in the Brillouin zone corresponding to the two $k_z$ values. The bands for $k_z=\pi/c$ are shifted so that the energy at the Dirac point matches that of $k_z=0$.

The monolayer graphene bands, shown in black, display the characteristic Dirac crossing at the K point. In AA graphite, the bands near the K/H points retain the linear dispersion associated with graphene, indicating that the low-energy $\pi$-electron structure remains dominated by the individual graphene layers. Away from the Dirac region, however, the bands split relative to monolayer graphene due to $k_z$-dependent inter-layer $\pi$--$\pi$ coupling, with bonding character at $k_z=0$ and antibonding character at $k_z=\pi/c$. Nevertheless, no gap opens at the Dirac point, confirming that inter-layer coupling modifies the band energies without breaking graphene's sublattice symmetry. \cite{CastroNeto2009_RMP_Graphene,Wallace1947_PR_Graphite,Lobato2011MultipleDirac}

Analogous plane-wave DFT calculations performed using Quantum ESPRESSO~\cite{Giannozzi2009QE,Giannozzi2017QE}, shown in Fig.~S3 of the SI, are in good agreement with the Wannier-first results in Fig.~\ref{fig_twisted_band}(a). This confirms that the Wannier-first framework captures the essential electronic features of AA graphite. As an additional comparison, Fig.~S4 presents the full AA graphite band structure alongside earlier tight-binding results in which the hopping parameters were obtained from \textit{ab initio} calculations.\cite{Lobato2011MultipleDirac} The overall band dispersions are in good agreement, including the $k_z$-dependent splitting of the graphene-derived Dirac bands while preserving their approximately linear dispersion near the $K/H$ points. The resulting overlap of electron and hole states is characteristic of the Dirac-like semi-metallic electronic structure of AA graphite.


We next introduce a rotational component into the symmetry operation to generate helically stacked AA (h-AA) graphite layers. In this generalized screw symmetry, each translation along $z$ is accompanied by a fixed rotation of the graphene sheet about the stacking axis. Figure~\ref{fig_twisted_band}(b) shows the resulting band structures for twist angles of $0^\circ$, $5^\circ$, $15^\circ$, and $25^\circ$. The $0^\circ$ case corresponds to the AA graphite structure discussed above.

For all twist angles considered, the Dirac crossing at the K point remains preserved, indicating that the low-energy electronic structure near the Fermi level is robust with respect to moderate rotational distortions. Away from the Dirac point, however, both occupied and unoccupied bands exhibit systematic changes with increasing twist angle. These variations reflect the modified overlap between $p_z$ orbitals in neighboring layers as the relative orientation of successive sheets changes. As the twist angle increases, 
subtle shifts and splittings appear across the band structure.

The total energies reported in Table~\ref{tab2} indicate that the twisted structures are energetically very close to the non-twisted AA configuration. For example, the energy difference between AA stacked  and helical (15$^\circ$) graphite is of the order of  $10^{-2}$~eV per unit cell. This weak dependence on twist angle is consistent with the relatively small magnitude of the inter-layer interaction compared with the strong in-plane covalent bonding. We note that no independent reference total-energy calculations were performed for the hAA graphite structures. 
The Wannier-first calculations for AA and hAA graphite were performed using the same WD.  For reference periodic calculations, a large supercell would be required for hAA graphite.

\section{Conclusions}
We have presented a periodic density functional framework in which Wannier-like functions are constructed directly from localized Gaussian basis functions without computing Bloch states during the SCF cycle. Building on the formalism of Pederson and Lin \cite{mark87}, the method generates a self-consistent set of localized Wannier-like functions within a finite Wannier domain (WD) and uses them to obtain the total charge density and total energy of 
extended systems. Electronic band structures are subsequently recovered in a post-processing step by computing matrix elements of the periodic Hamiltonian in a basis of Bloch orbitals.

The approach was validated for selected one-, two-, and three-dimensional systems. For linear and chiral $\mathrm{{-}C{\equiv}C{-}}$ and $\mathrm{{-}Li{-}F{-}}$ chains, the Wannier-first results reproduced independent periodic calculations performed with the Gaussian software package\cite{g16} and essentially the same basis sets, yielding band gaps in effectively exact agreement 
and total-energy differences on the order of $10^{-4}$~eV per unit cell. The method also captured symmetry-induced effects such as the lifting of band degeneracies in the helical chains and accurately reproduced energetics obtained from cluster-based extrapolations. 

The graphene benchmark demonstrates that the Wannier-first approach accurately reproduces localized orbitals, total energies, and the symmetry-protected Dirac cone of a prototypical two-dimensional material. The systematic convergence with respect to WD size confirms that the method provides a controllable and physically transparent real-space representation for two-dimensional periodic systems.

The formalism was further extended to three-dimensional layered systems by including translation or translation+rotation operations perpendicular to the graphene plane. Applications to AA graphite demonstrated that the Wannier-first framework accurately captures inter-layer $\pi$--$\pi$ coupling and the resulting $k_z$-dependent band splittings while preserving the symmetry-protected Dirac crossing. Comparisons with both plane-wave DFT and independent tight-binding calculations showed good overall agreement, confirming that the localized real-space formalism reproduces the 
electronic structure of AA graphite.  
The hAA graphite calculations demonstrated that arbitrary rotational symmetries can be treated using the same WD, eliminating the need for large supercells. The Dirac crossing remained preserved over a range of twist angles, while higher-energy bands exhibited systematic modifications arising from changes in inter-layer coupling.

The computational efficiency of the Wannier-first approach depends on the size of the WD used. This depends, in turn, on the spatial extent of the Gaussian basis functions. In particular, diffuse $s$- and $p$-type orbitals with small exponents produce slowly decaying tails, requiring substantially larger WDs to recover the periodic charge density to the desired numerical accuracy. This effect becomes especially important in three-dimensional systems, where the number of symmetry-related cells grows rapidly with domain size, increasing both memory requirements and computational cost. These limitations may be mitigated through recalibration of the Gaussian basis set by reducing the extent of the most diffuse functions and re-optimizing the orbital exponents to improve spatial localization while preserving electronic-structure accuracy. Creating an automated procedure for trimming the reach of the atomic basis functions to minimize the size of the WD needed for accurate calculations is a goal that is still to be reached. 

An important next step for the Wannier-first framework is using it to implement orbital-dependent exchange--correlation methods,\cite{RevModPhys.80.3} particularly SIC.\cite{perdew1981,heaton_1987,Stengel2008_PRB_77_155106,dabo_2010,Colonna2022_JCTC_Koopmans} Approximate density functionals suffer from self-interaction errors that can over-delocalize electron densities and lead to significant inaccuracies in systems containing stretched bonds, localized defect states, charge-transfer processes, and strongly correlated electrons.\cite{Perdew1985_IJQC_28_497,Sham1983_PRL_51_1888,Ruzsinszky2006_JCP_125_194112,Grafenstein2004_JCP_120_524,Patchkovskii2002_JCP_116_7806} 
It has been shown that the orbitals that minimize the Perdew-Zunger SIC total energy are localized, satisfying conditions known as the localization equations.\cite{mark_localization}  A Wannier-based representation therefore provides a natural framework for incorporating these corrections in periodic systems while preserving translational and screw symmetries. 

The motivation for the early work on the silicon crystal~\cite{mark87} was to obtain localized orbitals for calculating SIC. The (valence) Wannier functions associated with a given Si atom were expected to resemble $sp^{3}$ hybrid orbitals that would satisfy the localization equations\cite{mark_localization} amongst themselves. It was further expected that Wannier functions on different sites would satisfy the localization equations by symmetry. To realize these Wannier functions, a set of atom-centered functions of a given spin was chosen for one sub-lattice and the functions for the second sub-lattice were determined from the first by flipping the spin and inverting them through the bond axis (reflection would lead to a Lonsdaleite  form of silicon).  The resulting picture of two anti-ferromagnetically ordered lattices with vanishing total spin density appears to be a successful way to implement the Wannier-first algorithm as shown in Table S1. 

We anticipate using the Wannier-first method in conjunction with the Fermi-Lowdin-orbital (FLO) formalism\cite{mark2014,mark2015,Yang17,witha2018} regardless of whether standard DFT calculations or self-interaction corrected calculations are being performed.  
In such an extension, localized Fermi--L\"owdin orbitals (FLOs) would be constructed directly from the central-cell Wannier-like functions (and their translations) on each iteration, which should lead to even more localized Wannier representations.

 The results presented here establish the Wannier-first approach as an accurate and efficient alternative to conventional periodic methods for treating extended systems, including chiral and helical structures that are difficult to represent within standard translational periodicity. By combining a localized orbital representation with the explicit treatment of screw symmetry, the method enables compact simulations of chiral, helical, and rotationally modulated materials without requiring large supercells. The framework also provides a natural foundation for orbital-dependent exchange--correlation functionals, including Perdew--Zunger SIC,~\cite{perdew1981} and offers a promising route toward simulations of strongly correlated systems and large bio-molecular assemblies with helical symmetry, including DNA (deoxyribonucleic acid) and RNA (ribonucleic acid).

\section*{Supplementary Material}
See the Supporting Information for additional computational details, convergence tests, structural parameters, and supporting comparisons for the systems investigated in this work.

\begin{acknowledgments}
This work was supported by the U.S. Department of Energy,
Office of Science, Office of Basic Energy Sciences, as a part
of the Computational Chemical Sciences Program under Award
No. DE-SC0018331. The calculations were carried out at the
high-performance computing center (HPCC) of the Institute for
Cyber-Enabled Research (ICER) at Michigan State University using
resources owned by Central Michigan University.
\end{acknowledgments}

\noindent
\bibliographystyle{aipnum4-2}  
\bibliography{bibliography}

@article{Wannier90,
  author  = {Pizzi, Giovanni and Vitale, Valerio and Arita, Ryotaro and Bl{\"u}gel, Stefan and Freimuth, Frank and G{\'e}ranton, Gregory and Gibertini, Marco and Gresch, Dominik and Johnson, Corey and Koretsune, Takashi and others},
  title   = {Wannier90 as a community code: new features and applications},
  journal = {Journal of Physics: Condensed Matter},
  volume  = {32},
  number  = {16},
  pages   = {165902},
  year    = {2020},
  doi     = {10.1088/1361-648X/ab51ff}
}

@misc{g16,
author={M. J. Frisch and G. W. Trucks and H. B. Schlegel and G. E. Scuseria and M. A. Robb and J. R. Cheeseman and G. Scalmani and V. Barone and G. A. Petersson and H. Nakatsuji and X. Li and M. Caricato and A. V. Marenich and J. Bloino and B. G. Janesko and R. Gomperts and B. Mennucci and H. P. Hratchian and J. V. Ortiz and A. F. Izmaylov and J. L. Sonnenberg and D. Williams-Young and F. Ding and F. Lipparini and F. Egidi and J. Goings and B. Peng and A. Petrone and T. Henderson and D. Ranasinghe and V. G. Zakrzewski and J. Gao and N. Rega and G. Zheng and W. Liang and M. Hada and M. Ehara and K. Toyota and R. Fukuda and J. Hasegawa and M. Ishida and T. Nakajima and Y. Honda and O. Kitao and H. Nakai and T. Vreven and K. Throssell and Montgomery, {Jr.}, J. A. and J. E. Peralta and F. Ogliaro and M. J. Bearpark and J. J. Heyd and E. N. Brothers and K. N. Kudin and V. N. Staroverov and T. A. Keith and R. Kobayashi and J. Normand and K. Raghavachari and A. P. Rendell and J. C. Burant and S. S. Iyengar and J. Tomasi and M. Cossi and J. M. Millam and M. Klene and C. Adamo and R. Cammi and J. W. Ochterski and R. L. Martin and K. Morokuma and O. Farkas and J. B. Foresman and D. J. Fox},
title={Gaussian˜16 {R}evision {C}.01},
year={2016},
note={Gaussian Inc. Wallingford CT}
}

@article{Kresse1996_CMS_6_15,
  author  = {Kresse, G. and Furthm{\"u}ller, J.},
  title   = {Efficiency of ab-initio total energy calculations for metals and semiconductors using a plane-wave basis set},
  journal = {Computational Materials Science},
  volume  = {6},
  pages   = {15--50},
  year    = {1996},
  doi     = {10.1016/0927-0256(96)00008-0}
}

@article{mark87,
  title = {All-electron self-consistent variational method for Wannier-type functions: Applications to the silicon crystal},
  author = {Pederson, Mark R. and Lin, Chun C.},
  journal = {Phys. Rev. B},
  volume = {35},
  issue = {5},
  pages = {2273--2283},
  numpages = {0},
  year = {1987},
  month = {Feb},
  publisher = {American Physical Society},
  doi = {10.1103/PhysRevB.35.2273},
  url = {https://link.aps.org/doi/10.1103/PhysRevB.35.2273}
}

@article{kohn1965,
  title = {Self-Consistent Equations Including Exchange and Correlation Effects},
  author = {Kohn, W. and Sham, L. J.},
  journal = {Phys. Rev.},
  volume = {140},
  pages = {A1133--A1138},
  year = {1965}
}

@article{perdew1981,
  title = {Self-interaction correction to density-functional approximations for many-electron systems},
  author = {Perdew, J. P. and Zunger, A.},
  journal = {Phys. Rev. B},
  volume = {23},
  pages = {5048--5079},
  year = {1981}
}

@article{wannier1937,
  title = {The Structure of Electronic Excitation Levels in Insulating Crystals},
  author = {Wannier, G. H.},
  journal = {Phys. Rev.},
  volume = {52},
  pages = {191--197},
  year = {1937}
}

@article{marzari1997,
  title = {Maximally localized generalized Wannier functions for composite energy bands},
  author = {Marzari, N. and Vanderbilt, D.},
  journal = {Phys. Rev. B},
  volume = {56},
  pages = {12847--12865},
  year = {1997}
}

@article{nrlmol_pederson90,
  author  = {Pederson, Mark R. and Jackson, Kenneth A.},
  title   = {Variational mesh for quantum-mechanical simulations},
  journal = {Physical Review B},
  volume  = {41},
  number  = {11},
  pages   = {7453--7461},
  year    = {1990},
  doi     = {10.1103/PhysRevB.41.7453}
}

@article{nrlmol_jack90,
  author  = {Jackson, Kenneth A. and Pederson, Mark R.},
  title   = {Accurate forces in density-functional calculations},
  journal = {Physical Review B},
  volume  = {42},
  number  = {6},
  pages   = {3276--3281},
  year    = {1990},
  doi     = {10.1103/PhysRevB.42.3276}
}

@article{nrlmol_porezag99,
  author  = {Porezag, Dirk and Pederson, Mark R.},
  title   = {Optimization of Gaussian basis sets for density-functional calculations},
  journal = {Physical Review A},
  volume  = {60},
  number  = {4},
  pages   = {2840--2847},
  year    = {1999},
  doi     = {10.1103/PhysRevA.60.2840}
}

@article{nrlmol_pederson2000,
  author  = {Pederson, Mark R. and Heaton, Richard A. and Lin, Ching C.},
  title   = {Local basis sets and self-interaction corrections in density-functional theory},
  journal = {physica status solidi (b)},
  volume  = {217},
  number  = {1},
  pages   = {197--218},
  year    = {2000},
  doi     = {10.1002/(SICI)1521-3951(200001)217:1<197::AID-PSSB197>3.0.CO;2-V}
}

@article{Perdew1992,
  title     = {Accurate and simple analytic representation of the electron-gas correlation energy},
  author    = {Perdew, John P. and Wang, Yue},
  journal   = {Physical Review B},
  volume    = {45},
  number    = {23},
  pages     = {13244--13249},
  year      = {1992},
  month     = jun,
  publisher = {American Physical Society},
  doi       = {10.1103/PhysRevB.45.13244}
}

@article{Colonna2022_JCTC_Koopmans,
  author  = {Colonna, Nicola and De Gennaro, Riccardo and Linscott, Edward and Marzari, Nicola},
  title   = {Koopmans Spectral Functionals in Periodic Boundary Conditions},
  journal = {Journal of Chemical Theory and Computation},
  volume  = {18},
  number  = {9},
  year    = {2022},
  publisher = {American Chemical Society}
}

@article{Stengel2008_PRB_77_155106,
  author  = {Stengel, Massimiliano and Spaldin, Nicola A.},
  title   = {Self-interaction correction with Wannier functions},
  journal = {Physical Review B},
  volume  = {77},
  pages   = {155106},
  year    = {2008},
  doi     = {10.1103/PhysRevB.77.155106}
}

@article{Marzari2012_RMP_84_1419,
  author  = {Marzari, Nicola and Mostofi, Arash A. and Yates, Jonathan R. and Souza, Ivo and Vanderbilt, David},
  title   = {Maximally localized Wannier functions: Theory and applications},
  journal = {Reviews of Modern Physics},
  volume  = {84},
  pages   = {1419--1475},
  year    = {2012},
  doi     = {10.1103/RevModPhys.84.1419}
}

@article{Bloom2024_ChemRev_CISS,
  author  = {Bloom, Brian P. and Paltiel, Yossi and Naaman, Ron and Waldeck, David H.},
  title   = {Chiral Induced Spin Selectivity},
  journal = {Chemical Reviews},
  volume  = {124},
  number  = {4},
  pages   = {1950--2040},
  year    = {2024},
  doi     = {10.1021/acs.chemrev.3c00635}
}

@article{Evers2022_AdvMater_CISS,
  author  = {Evers, Ferdinand and Aharony, Amnon and Bar-Gill, Nir and Entin-Wohlman, Ora and Hedeg{\aa}rd, Per and Hod, Oded and Jelinek, Pavel and Kamieniarz, Grzegorz and Lemeshko, Mikhail and Michaeli, Karen and Mujica, Vladimiro and Naaman, Ron and Paltiel, Yossi and Refaely-Abramson, Sivan and Tal, Oren and Thijssen, Jos and Thoss, Michael and van Ruitenbeek, Jan M. and Venkataraman, Latha and Waldeck, David H. and Yan, Binghai and Kronik, Leeor},
  title   = {Theory of Chirality Induced Spin Selectivity: Progress and Challenges},
  journal = {Advanced Materials},
  volume  = {34},
  number  = {13},
  pages   = {2106629},
  year    = {2022},
  doi     = {10.1002/adma.202106629}
}

@article{Han2024_JACS_COF_CISS,
  author  = {Han, Xing and Jiang, Chao and Hou, Bang and Liu, Yan and Cui, Yong},
  title   = {Covalent Organic Frameworks with Tunable Chirality for Chiral-Induced Spin Selectivity},
  journal = {Journal of the American Chemical Society},
  volume  = {146},
  number  = {10},
  pages   = {7045--7052},
  year    = {2024},
  doi     = {10.1021/jacs.3c13283}
}

@article{Naaman2022_AnnuRevBiophys_51_99,
  author  = {Naaman, Ron and Paltiel, Yossi and Waldeck, David H.},
  title   = {Chiral Induced Spin Selectivity and Its Implications for Biological Functions},
  journal = {Annual Review of Biophysics},
  volume  = {51},
  pages   = {99--114},
  year    = {2022},
  doi     = {10.1146/annurev-biophys-083021-070400}
}

@article{Wang2025_NanoLett_ChiralMagnet,
  author  = {Wang, Zhongxuan and Xie, Ti and Fang, Zhenyao and Zhang, Jun and Gong, Cheng and Yan, Qimin and Ren, Shenqiang},
  title   = {Chiral Molecular Magnet Superstructures with Light Control},
  journal = {Nano Letters},
  volume  = {25},
  number  = {6},
  year    = {2025},
  doi     = {10.1021/acs.nanolett.4c05606}
}

@article{Getahun2023_APL_122_241903,
  author  = {Getahun, Yohannes W. and Manciu, Felicia S. and Pederson, Mark R. and El-Gendy, Ahmed A.},
  title   = {Room temperature colossal superparamagnetic order in aminoferrocene--graphene molecular magnets},
  journal = {Applied Physics Letters},
  volume  = {122},
  pages   = {241903},
  year    = {2023},
  doi     = {10.1063/5.0153212}
}

@article{Perdew1985_IJQC_28_497,
  author  = {Perdew, J. P.},
  title   = {Density functional theory and the band gap problem},
  journal = {International Journal of Quantum Chemistry},
  volume  = {28},
  pages   = {497--523},
  year    = {1985},
  doi     = {10.1002/qua.560280846}
}

@article{Sham1983_PRL_51_1888,
  author  = {Sham, L. J. and Schl{\"u}ter, M.},
  title   = {Density-functional theory of the energy gap},
  journal = {Physical Review Letters},
  volume  = {51},
  pages   = {1888--1891},
  year    = {1983},
  doi     = {10.1103/PhysRevLett.51.1888}
}

@article{Patchkovskii2002_JCP_116_7806,
  author  = {Patchkovskii, S. and Ziegler, T.},
  title   = {Improving ``difficult'' reaction barriers with self-interaction corrected density functional theory},
  journal = {The Journal of Chemical Physics},
  volume  = {116},
  pages   = {7806--7813},
  year    = {2002},
  doi     = {10.1063/1.1468640}
}

@article{Grafenstein2004_JCP_120_524,
  author  = {Gr{\"a}fenstein, J. and Kraka, E. and Cremer, D.},
  title   = {The impact of the self-interaction error on the density functional theory description of dissociating radical cations: Ionic and covalent dissociation limits},
  journal = {The Journal of Chemical Physics},
  volume  = {120},
  pages   = {524--539},
  year    = {2004},
  doi     = {10.1063/1.1630017}
}

@article{Ruzsinszky2006_JCP_125_194112,
  author  = {Ruzsinszky, A. and Perdew, J. P. and Csonka, G. I. and Vydrov, O. A. and Scuseria, G. E.},
  title   = {Spurious fractional charge on dissociated atoms: Pervasive and resilient self-interaction error of common density functionals},
  journal = {The Journal of Chemical Physics},
  volume  = {125},
  pages   = {194112},
  year    = {2006},
  doi     = {10.1063/1.2387954}
}

@article{Lobato2011MultipleDirac,
  title = {Multiple Dirac particles in AA-stacked graphite and multilayers of graphene},
  author = {Lobato, I. and Partoens, B.},
  journal = {Physical Review B},
  volume = {83},
  number = {16},
  pages = {165429},
  year = {2011},
  month = apr,
  publisher = {American Physical Society},
  doi = {10.1103/PhysRevB.83.165429}
}

@article{CastroNeto2009_RMP_Graphene,
  author  = {Castro Neto, A. H. and Guinea, F. and Peres, N. M. R. and Novoselov, K. S. and Geim, A. K.},
  title   = {The electronic properties of graphene},
  journal = {Reviews of Modern Physics},
  volume  = {81},
  number  = {1},
  pages   = {109--162},
  year    = {2009},
  doi     = {10.1103/RevModPhys.81.109}
}

@article{Wallace1947_PR_Graphite,
  author  = {Wallace, P. R.},
  title   = {The Band Theory of Graphite},
  journal = {Physical Review},
  volume  = {71},
  number  = {9},
  pages   = {622--634},
  year    = {1947},
  doi     = {10.1103/PhysRev.71.622}
}

@article{Giannozzi2009QE,
  title     = {QUANTUM ESPRESSO: a modular and open-source software project for quantum simulations of materials},
  author    = {Giannozzi, Paolo and Baroni, Stefano and Bonini, Nicola and Calandra, Matteo and Car, Roberto and Cavazzoni, Carlo and Ceresoli, Davide and Chiarotti, Guido L. and Cococcioni, Matteo and Dabo, Ismaila and Dal Corso, Andrea and de Gironcoli, Stefano and Fabris, Stefano and Fratesi, Guido and Gebauer, Ralph and Gerstmann, Uwe and Gougoussis, Christos and Kokalj, Anton and Lazzeri, Michele and Martin-Samos, Layla and Marzari, Nicola and Mauri, Francesco and Mazzarello, Riccardo and Paolini, Stefano and Pasquarello, Alfredo and Paulatto, Lorenzo and Sbraccia, Carlo and Scandolo, Sandro and Sclauzero, Gabriele and Seitsonen, Ari P. and Smogunov, Alexander and Umari, Paolo and Wentzcovitch, Renata M.},
  journal   = {Journal of Physics: Condensed Matter},
  volume    = {21},
  number    = {39},
  pages     = {395502},
  year      = {2009},
  doi       = {10.1088/0953-8984/21/39/395502}
}

@article{Giannozzi2017QE,
  title     = {Advanced capabilities for materials modelling with Quantum ESPRESSO},
  author    = {Giannozzi, Paolo and Andreussi, Oliviero and Brumme, Thomas and Bunau, Olivier and Buongiorno Nardelli, Marco and Calandra, Matteo and Car, Roberto and Cavazzoni, Carlo and Ceresoli, Davide and Cococcioni, Matteo and Colonna, Nicola and Carnimeo, Ivan and Dal Corso, Andrea and de Gironcoli, Stefano and Delugas, Pietro and DiStasio Jr., Robert A. and Ferretti, Andrea and Floris, Andrea and Fratesi, Guido and Fugallo, Giorgia and Gebauer, Ralph and Gerstmann, Uwe and Giustino, Feliciano and Gorni, Tommaso and Jia, Jiayu and Kawamura, Mitsuaki and Ko, Hyungjun-Yong and Kokalj, Anton and K{\"u}{\c c}{\"u}kbenli, Emine and Lazzeri, Michele and Marsili, Matteo and Marzari, Nicola and Mauri, Francesco and Nguyen, Nguyen L. and Nguyen, Hung-Vu and Otero-de-la-Roza, Alberto and Paulatto, Lorenzo and Ponc{\'e}, Samuel and Rocca, Dario and Sabatini, Riccardo and Santra, Biswajit and Schlipf, Martin and Seitsonen, Ari P. and Smogunov, Alexander and Timrov, Iurii and Thonhauser, Timo and Umari, Paolo and Vast, Nathalie and Wu, Xifan and Baroni, Stefano},
  journal   = {Journal of Physics: Condensed Matter},
  volume    = {29},
  number    = {46},
  pages     = {465901},
  year      = {2017},
  doi       = {10.1088/1361-648X/aa8f79}
}

@article{RevModPhys.80.3,
  title = {Orbital-dependent density functionals: Theory and applications},
  author = {K\"ummel, Stephan and Kronik, Leeor},
  journal = {Rev. Mod. Phys.},
  volume = {80},
  issue = {1},
  pages = {3--60},
  numpages = {0},
  year = {2008},
  month = {Jan},
  publisher = {American Physical Society},
  doi = {10.1103/RevModPhys.80.3},
  url = {https://link.aps.org/doi/10.1103/RevModPhys.80.3}
}

@article{mark2014,
    author = {Pederson, Mark R. and Ruzsinszky, Adrienn and Perdew, John P.},
    title = {Communication: Self-interaction correction with unitary invariance in density functional theory},
    journal = {The Journal of Chemical Physics},
    volume = {140},
    number = {12},
    pages = {121103},
    year = {2014},
    month = {03},
    issn = {0021-9606},
    doi = {10.1063/1.4869581},
}

@article{mark2015,
author = {Pederson,Mark R. },
title = {Fermi orbital derivatives in self-interaction corrected density functional theory: Applications to closed shell atoms},
journal = {J. Chem. Phys.},
volume = {142},
number = {6},
pages = {064112},
year = {2015},
doi = {10.1063/1.4907592},
}

@article{Yang17,
  title = "{Full self-consistency in the Fermi-orbital self-interaction correction}",
  author = {Yang, Zeng-hui and Pederson, Mark R. and Perdew, John P.},
  journal = {Phys. Rev. A},
  volume = {95},
  issue = {5},
  pages = {052505},
  numpages = {8},
  year = {2017},
  month = {May},
  publisher = {American Physical Society},
  doi = {10.1103/PhysRevA.95.052505}
}

@article{witha2018,
    author = {Withanage, K.P.K and Trepte, K. and Peralta, J. E. and Baruah, T. and Zope, R. R. and Jackson, K. A.},
    title = {On the Question of the Total Energy in the Fermi–Löwdin Orbital Self-Interaction Correction Method},
    journal = {J. Chem. Theor. Comput.},
    volume = {14},
    pages = {4122},
    year = {2018}
}

@article{mark_localization,
    author = {Pederson, M. R. and Lin, C. C.},
    title = {Localized and canonical atomic orbitals in self-interaction corrected local density functional approximation},
    journal = {J. Chem. Phys.},
    volume = {88},
    pages = {1807},
    year = {1988}
}

@article{heaton_1987,
    author = {Heaton, R. A. and Pederson, M. R. and Lin, C. C.},
    title = {A new density functional for fractionally occupied orbital systems with application to ionization and transition energies},
    journal = {J. Chem. Phys.},
    volume = {86},
    pages = {258},
    year = {1987}
}

@article{dabo_2010,
    author = {Dabo, I. and Ferretti, A. and Poilvert, N. and Li, Y. and Marzari, N. and Cococcioni, M.},
    title = {Koopmans' condition for density-functional theory},
    journal = {Phys. Rev. B},
    volume = {82},
    pages = {115121},
    year = {2010}
}

@article{Alase2017GeneralizedBloch,
  author  = {Alase, Abhijeet and Cobanera, Emilio and Ortiz, Gerardo and Viola, Lorenza},
  title   = {Generalization of Bloch's theorem for arbitrary boundary conditions: Theory},
  journal = {Physical Review B},
  volume  = {96},
  number  = {19},
  pages   = {195133},
  year    = {2017},
  doi     = {10.1103/PhysRevB.96.195133}
}

@article{Cobanera2018GeneralizedBloch,
  author  = {Cobanera, Emilio and Alase, Abhijeet and Ortiz, Gerardo and Viola, Lorenza},
  title   = {Generalization of Bloch's theorem for arbitrary boundary conditions: Interfaces and topological surface band structure},
  journal = {Physical Review B},
  volume  = {98},
  number  = {24},
  pages   = {245423},
  year    = {2018},
  doi     = {10.1103/PhysRevB.98.245423}
}

@article{Dobardzic2015GeneralizedBloch,
  author  = {Dobard{\v{z}}i{\'c}, E. and Dimitrijevi{\'c}, M. and Milovanovi{\'c}, M. V.},
  title   = {Generalized Bloch theorem and topological characterization},
  journal = {Physical Review B},
  volume  = {91},
  number  = {12},
  pages   = {125424},
  year    = {2015},
  doi     = {10.1103/PhysRevB.91.125424}
}

@article{Baruah2009,
  author    = {Baruah, Tunna and Pederson, Mark R.},
  title     = {DFT Calculations on Charge-Transfer States of a Carotenoid-Porphyrin-$\mathrm{C}_{60}$ Molecular Triad},
  journal   = {Journal of Chemical Theory and Computation},
  volume    = {5},
  number    = {4},
  pages     = {834--843},
  year      = {2009},
  publisher = {American Chemical Society},
  doi       = {10.1021/ct900024f},
  url       = {https://doi.org/10.1021/ct900024f}
}

@article{Baruah2006,
  author    = {Baruah, Tunna and Pederson, Mark R.},
  title     = {Density functional study on a light-harvesting carotenoid-porphyrin-$\mathrm{C}_{60}$ molecular triad},
  journal   = {The Journal of Chemical Physics},
  volume    = {125},
  number    = {16},
  pages     = {164706},
  year      = {2006},
  publisher = {AIP Publishing},
  doi       = {10.1063/1.2360265},
  url       = {https://doi.org/10.1063/1.2360265}
}

@article{Mustafa2015OPFMWannier,
  author  = {Mustafa, Jamal I. and Coh, Sinisa and Cohen, Marvin L. and Louie, Steven G.},
  title   = {Automated construction of maximally localized {Wannier} functions: Optimized projection functions method},
  journal = {Physical Review B},
  volume  = {92},
  issue   = {16},
  pages   = {165134},
  year    = {2015},
  doi     = {10.1103/PhysRevB.92.165134},
  url     = {https://doi.org/10.1103/PhysRevB.92.165134}
}

@article{Mustafa2016TopologicalWannier,
  author  = {Mustafa, Jamal I. and Coh, Sinisa and Cohen, Marvin L. and Louie, Steven G.},
  title   = {Automated construction of maximally localized {Wannier} functions for bands with nontrivial topology},
  journal = {Physical Review B},
  volume  = {94},
  issue   = {12},
  pages   = {125151},
  year    = {2016},
  doi     = {10.1103/PhysRevB.94.125151},
  url     = {https://doi.org/10.1103/PhysRevB.94.125151}
}

\end{document}